\documentclass[jap,aip,amsmath,amssymb,reprint,floatfix,longbibliography]{revtex4-2}

\usepackage[utf8]{inputenc}
\usepackage[T1]{fontenc}

\usepackage{amsfonts}
\usepackage{stmaryrd}
\usepackage{graphicx}
\usepackage{siunitx}
\usepackage{xcolor}
\usepackage{physics} 
\usepackage{multirow} 
\usepackage{ulem} 
\usepackage[colorlinks=true,citecolor=blue,urlcolor=red]{hyperref}
\usepackage{orcidlink}

\newcommand{\CP}{\mathrm{B}}

\newcommand{\eff}{\mathrm{eff}}

\begin{document}

\title{Thermal noise calibration of functionalized cantilevers for force microscopy: effects of the colloidal probe position.}

\author{Aubin Archambault\,\orcidlink{0009-0002-3373-357X}}
\affiliation{Univ Lyon, ENS de Lyon, CNRS, Laboratoire de Physique, F-69342 Lyon, France}
\author{Caroline Crauste-Thibierge\,\orcidlink{0000-0001-5502-0445}}
\affiliation{Univ Lyon, ENS de Lyon, CNRS, Laboratoire de Physique, F-69342 Lyon, France}
\author{Ludovic Bellon\,\orcidlink{0000-0002-2499-8106}}
\email{ludovic.bellon@ens-lyon.fr}
\affiliation{Univ Lyon, ENS de Lyon, CNRS, Laboratoire de Physique, F-69342 Lyon, France}

\date{\today}

\begin{abstract}
Colloidal probes are often used in force microscopy when the geometry of the tip-sample interaction should be well controlled. Their calibration requires the understanding of their mechanical response, which is very sensitive to the details of the force sensor consisting of a cantilever and the attached colloid. We present analytical models to describe the dynamics of the cantilever and its load positioned anywhere along its length. The thermal noise calibration of such probes is then studied from a practical point of view, leading to correction coefficients that can be applied in standard force microscope calibration routines. Experimental measurements of resonance frequencies and thermal noise profiles of raw and loaded cantilevers demonstrate the validity of the approach.
\end{abstract}

\maketitle 

\section{Introduction}

Atomic Force Microscopy (AFM) is nowadays a routine technology in many laboratories, from material science to biophysics~\cite{Giessibl-2003,Butt-2005,Kuznetsova-2007,Muller-2008}. This scanning microscopy is based on the interaction between a local probe and a sample. The force is recorded by monitoring the deflection of a cantilever supporting the local probe, designated as the tip. In many applications, the control of the tip shape is beneficial to an enhanced reproducibility and to reach quantitative measurements~\cite{Maali-2008,Laurent-2012,Kim-2015,Chighizola-2020,Eskhan-2022,Holuigue-2022,Kim-2022,Sharma-2022,Xu-2022,Zielinski-2022,Zimron-Politi-2023,Lorenc-2023}. Colloidal probes trade the tip sharpness and associated spatial resolution for a better knowledge of the probe geometry and a better interpretation of the force signal. They are manufactured by attaching to the cantilever a bead with a diameter ranging from a few to over a hundred micrometers. To fully exploit the better knowledge of the interaction geometry and reach quantitative measurements, the probe itself and its mechanical response should be well characterized and calibrated. 

One key question is the calibration of the force sensitivity~\cite{Burnham-2003,Butt-2005}: how do we translate the measured deflection of the cantilever into a force of interaction? To answer this question, we need to understand how a force applied locally at the tip deflects the cantilever, and how this deflection is measured. We therefore need a description of the cantilever shape, and of a calibration technique to apply a known force and read the corresponding deflection. The latter part is often performed using a thermal noise calibration\cite{Butt-1995}: in equilibrium at room temperature $T$, the environnement exerts a random force whose statistical properties are described by the fluctuation-dissipation theorem. An harmonic oscillator of stiffness $k$ for example presents a mean square displacement $\langle Z^2 \rangle = k_B T / k$, where $k_B$ is the Boltzmann constant. Measuring $\langle Z^2 \rangle$ therefore leads to the value of $k$. The complexity of the problem increases for a cantilever, which is a spatially extended object, whose deflection is generally measured locally with some optical technique. For small tips, an Euler-Bernoulli description of the cantilever by a clamped-free mechanical beam is well suited, and leads to the established thermal noise calibration of the probe sensitivity\cite{Butt-1995}. The introduction of a relatively heavy and large load close to its free end modifies significantly the dynamical behavior to the force probe, and should be addressed. Some approaches rely for example on finite element analysis for the determination of calibration parameters~\cite{Rodriguez-Ramos-2021}. In Refs.~\onlinecite{Li-2008,Allen-2009,Laurent-2013,Chighizola-2020}, extensions of the analytical approach of the Euler Bernoulli model~\cite{Oguamanam-2003} to beads glued at the free end of the cantilever are worked out.

In this article, we further refine these last approaches to take into account more precisely the geometry of the probe. Indeed, previous approaches are restricted to a colloidal bead attached precisely at the extremity of the cantilever, which is an idealization of the experimental situation. To be more faithful to real world probes, we add here the possibility of a setback with respect to this free end (section \ref{sec:analytical_solution_SPC}), or of a possible rigidification of the cantilever free end due to the gluing process (section \ref{sec:analytical_solution_Endload}). The questions are then to understand what is the shape of the cantilever's deflection, and how should one handle the thermal noise calibration (section \ref{sec:thermalnoisecalib}). Once the theoretical stage set, we demonstrate the validity of the models introduced with experiments on three different cantilevers (section \ref{sec:exp}). The added loads are characterized by studying how the resonance frequencies of the lever are modified by the functionalisation, and the resonance mode shape are compared to the prediction of the two models. A conclusion finally summarizes the key points of the article, and how they can be applied to practical cases, with functional scripts to compute the relevant calibration coefficients for experimentalists~\cite{Archambault-Scripts-2023}.

\section{Analytical solution, single point contact}\label{sec:analytical_solution_SPC}

\begin{figure}[ht]
\begin{center}
\includegraphics{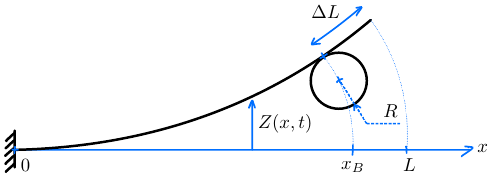}
\end{center}
\caption{Sketch of an AFM cantilever of length $L$ loaded with a colloïdal particle of radius $R$ at an offset $\Delta L$ from the free end. The bead is supposed here to have a single point of contact with the cantilever, where it can exert a force and a torque. $Z(x,t)$ describes the deflection along the cantilever.}
\label{fig:sketchPunctual} 
\end{figure}

We consider a rectangular cantilever of length $L$, uniform width $W$ and thickness $H$, functionalized with a colloidal particle of radius $R$, as sketch in Fig.~\ref{fig:sketchPunctual}. This sphere is supposed to be glued on the symmetry axis of the cantilever, at a distance $\Delta L$ from the cantilever free end. We neglect in a first approximation the effect of rigidification due to the gluing process: the contact between the bead and the cantilever is supposed to be punctual, and the bead infinitely rigid. It will influence the cantilever motion by adding a local force (due to its inertia in translation) and torque (inertia in rotation). We work under the hypothesis of an elongated beam: the cantilever is supposed to be much longer than its transverse dimensions ($L \gg W,H$). We therefore use the Euler Bernoulli equation for the deflection $Z(x,t)$ along the cantilever length:
\begin{equation} \label{eq:EB}
\frac{EI}{L^4}\partial_x^4 Z(x,t) + \mu A \partial_t^2 Z(x,t) = 0,
\end{equation}
where $I=WH^3/12$ is the second moment of inertia of the beam, $A=WH$ its cross section area, $E$ the material Young's modulus, $\mu$ its density, $t$ the time and $x$ the position along the cantilever normalised by $L$. In absence of external forces, four boundary conditions (BC) shall be applied:
\begin{subequations}
\label{eq:BC}
\begin{align}
Z(0,t) &= \partial_x Z(0,t) = 0, \label{eq:BC0} \\
\partial_x^2 Z(1,t) &= \partial_x^3 Z(1,t) = 0. \label{eq:BC1}
\end{align}
\end{subequations}
The first two BC express the clamping at $x=0$, the last two hold for the absence of torque and force at the free end $x=1$. The colloidal particle adds a pair of extra conditions, as by inertia it adds a force $-m_\CP \partial_t^2 Z(x_\CP,t)$ and a torque $I_\CP/L \partial_t^2 \partial_x Z(x_\CP,t)$ at $x_\CP=1-\Delta L/L$, with $m_\CP$ and $I_\CP$ the bead's mass and moment of inertia (at the contact point with the cantilever). It translates in the following jump conditions:
\begin{subequations}
\label{eq:jumpxCP}
\begin{align}
\frac{EI}{L^2} \big\llbracket \partial_x^2 Z\big\rrbracket (x_\CP,t) & = \frac{I_\CP}{L} \partial_t^2 \partial_x Z(x_\CP,t), \label{eq:jumpxCP2} \\
\frac{EI}{L^3} \big\llbracket\partial_x^3 Z\big\rrbracket (x_\CP,t) & = -m_\CP \partial_t^2 Z(x_\CP,t), \label{eq:jumpxCP3} 
\end{align}
\end{subequations}
where $\big\llbracket . \big\rrbracket(x_\CP)$ stand for the jump of the quantity when crossing $x_\CP$: $\big\llbracket Z \big\rrbracket (x_\CP) = \lim_{\epsilon\rightarrow 0} [Z (x_\CP+\epsilon) - Z (x_\CP-\epsilon)]$. Note that the cantilever and its slope are continuous at $x_\CP$, so that
\begin{equation}
\label{eq:contxCP}
\big\llbracket Z\big\rrbracket (x_\CP,t) = \big\llbracket\partial_x Z\big\rrbracket (x_\CP,t) = 0.
\end{equation}

Let us solve this problem by separating space and time: with $Z(x,t)=z(x)e^{i \omega t}$, Eq.~\ref{eq:EB} can be rewritten as
\begin{equation} \label{eq:EB2}
z^{(4)}(x) = \alpha^4 z(x),
\end{equation}
where superscript $^{(n)}$ stands for the $n^\mathrm{th}$ derivative, and $\alpha$ is given by the dispersion relation:
\begin{equation} \label{eq:dispersion}
\alpha^4 = \frac{\mu A L^4}{E I} \omega^2.
\end{equation}
For the boundary conditions, Eqs.~\eqref{eq:BC} and \eqref{eq:contxCP} apply directly replacing $Z$ by $z$, and the jump condition can be rewritten as:
\begin{subequations}
\label{eq:jumpxCPz}
\begin{align}
\big\llbracket z''\big\rrbracket (x_\CP) & = - \tilde{m} \tilde{\rho}^2 \alpha^4 z'(x_\CP), \label{eq:jumpxCP2z} \\
\big\llbracket z^{(3)}\big\rrbracket (x_\CP) & = \tilde{m} \alpha^4 z(x_\CP), \label{eq:jumpxCP3z} 
\end{align}
\end{subequations}
where $\tilde{m} = m_\CP / m_c$ is the bead's mass normalised by the cantilever's one $m_c=\mu A L$, and $\tilde{\rho} = \sqrt{I_\CP/(m_\CP L^2)}$ is the normalised giration radius of the bead (including both the intrinsic moment of inertia of the load and its distance to the cantilever surface).

Between $0$ and $x_\CP$ and between $x_\CP$ and $1$, we thus need to solve two boundary value problems of order 4, implying 8 coefficients, we thus need to use 8 boundary conditions to solve it: Eqs.~\eqref{eq:BC}, \eqref{eq:contxCP} and \eqref{eq:jumpxCPz}. Fortunately, using an analytical shooting method, the solution can be found with less complexity. We introduce the function $\psi_c(x)$ and $\psi_s(x)$ defined by
\begin{subequations} \label{eq:psics}
\begin{align}
\psi_c(x) & = H(x) \frac{\cosh \alpha x - \cos \alpha x}{2 \alpha^2}, \\
\psi_s(x) & = H(x) \frac{\sinh \alpha x - \sin \alpha x}{2 \alpha^3},
\end{align}
\end{subequations}
with $H(x)$ the Heaviside function ($0$ if $x<0$, $1$ otherwise). Both are solutions of Eq.~\eqref{eq:EB2}, and verify the BC of Eqs.~\eqref{eq:BC0} at $x=0$. Moreover, they are designed such that $\psi_c''(0)=1$, $\psi_c^{(3)}(0)=0$, $\psi_s''(0)=0$, $\psi_s^{(3)}(0)=1$. Both functions $\psi_c(x)$ and $\psi_s(x)$, solution of the equation between $0$ and $x_\CP$, must be completed above $x_\CP$ to meet the jump conditions \eqref{eq:jumpxCPz} and continuity ones \eqref{eq:contxCP}. This is conveniently done by adding two terms in $\psi_c(x-x_\CP)$ and $\psi_s(x-x_\CP)$ above $x_\CP$, with coefficients corresponding to the jump of the second derivative for $\psi_c$ (Eq.~\ref{eq:jumpxCP2z}) and of the third derivative for $\psi_s$ (Eq.~\ref{eq:jumpxCP3z}):
\begin{subequations} \label{eq:phics}
\begin{align}
\begin{split}
\phi_c(x) = \psi_c(x) + \tilde m \alpha^4 \big[ & \psi_c(x_\CP) \psi_s(x-x_\CP) \\
&- \tilde{\rho}^2 \psi_c'(x_\CP) \psi_c(x-x_\CP)\big],
\end{split} \\
\begin{split}
\phi_s(x) = \psi_s(x) + \tilde m \alpha^4 \big[ &\psi_s(x_\CP) \psi_s(x-x_\CP) \\
&- \tilde{\rho}^2 \psi_s'(x_\CP) \psi_c(x-x_\CP)\big].
\end{split}
\end{align}
\end{subequations}
Those two functions $\phi_c$ and $\phi_s$ now fulfil the BC at $x=0$ and the jump conditions at $x=x_\CP$. We are only left to take a combination of both and write the BC at $x=1$:
\begin{subequations} \label{eq:combinephics}
\begin{align}
z(x) &= a \phi_c(x) + b \phi_s(x), \\
z''(1) &= a \phi_c''(1) + b \phi_s''(1) = 0, \\
z^{(3)}(1) &= a \phi_c^{(3)}(1) + b \phi_s^{(3)}(1) = 0.
\end{align}
\end{subequations}
Since $(a,b)\neq(0,0)$, the following condition must be fulfilled:
\begin{equation}
\phi_c''(1) \phi_s^{(3)}(1) = \phi_s''(1) \phi_c^{(3)}(1).
\end{equation}
This implicit equation on $\alpha$ prescribes the countable set of spatial eigenvalues $\alpha_n(\tilde{m},\tilde{\rho},x_\CP)$ of the resonant modes of the clamped cantilever functionalized by the colloidal particle. Each $\alpha_n$ corresponds to a resonance angular frequency $\omega_n$ given by Eq.~\eqref{eq:dispersion}.

Using Eqs.~\eqref{eq:combinephics}, the eigenmodes $\psi_n$ are for example given by:
\begin{equation} \label{eq:psin}
\psi_n (x) = \phi_c(x) - R_n \phi_s(x),
\end{equation}
where $R_n=\phi_c''(1)/\phi_s''(1)$ is an implicit function of $\alpha_n$. To define an orthonormal base of eigenmodes, we need to introduce the scalar product. For 2 eigenmodes $\psi_n$ and $\psi_p$ associated with eigenvalues $\alpha_n$ and $\alpha_p$, let us compute
\begin{equation}
\begin{split}
\alpha_n^4 \int_0^1 \psi_n (x) \psi_p (x) \dd x = \int_0^1 \psi_n^{(4)}(x) \psi_p (x) \dd x = \\
\lim_{\epsilon\rightarrow0} \int_0^{x_\CP-\epsilon} \psi_n^{(4)}(x) \psi_p (x) \dd x + \int_{x_\CP+\epsilon}^1 \psi_n^{(4)}(x) \psi_p (x) \dd x.
\end{split}
\end{equation}
By carefully including the jump conditions in $x=x_\CP$, after four integrations by parts we get:
\begin{equation}
\begin{split}
(\alpha_n^4-\alpha_p^4)\bigg( \int_0^1 \psi_n (x) \psi_p (x) \dd x + \tilde{m} \psi_n(x_\CP) \psi_p(x_\CP) \\
+ \tilde{m} \tilde{\rho}^2 \psi_n'(x_\CP) \psi_p'(x_\CP)\bigg)=0.
\end{split}
\end{equation}
When $\alpha_n\neq \alpha_p$, the right parenthesis must be zero and is chosen to define the scalar product $\langle\psi_n.\psi_p\rangle$. We can eventually define the orthonormal base $\phi_n$ of normal modes by
\begin{equation} \label{eq:phin}
\phi_n(x)=\frac{1}{\sqrt{\langle\psi_n^2\rangle}} \psi_n(x),
\end{equation}
where the normalisation factor is the norm of $\psi_n$, i.e.
\begin{equation}
\langle\psi_n^2\rangle = \int_0^1 \psi_n^2 (x) \dd x + \tilde{m} \psi_n^2(x_\CP) + \tilde{m} \tilde{\rho}^2 \big[\psi_n'(x_\CP)\big]^2.
\end{equation}
This last equation can be understood as the repartition of kinetic energy of the normal mode between the elastic beam (first term), the translation of the bead (second term), and its rotation (last term)~\cite{Laurent-2013}. From Eqs.~\eqref{eq:psics}, \eqref{eq:phics}, \eqref{eq:psin} and \eqref{eq:phin}, the explicit form of $\phi_n(x)$ can be written, but its long and complex expression is of little interest.

\begin{figure}[ht]
\begin{center}
\includegraphics{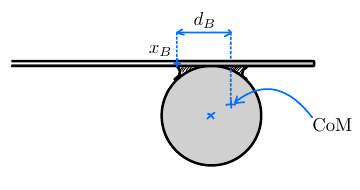}
\end{center}
\caption{Sketch of the cantilever in the approximation of a rigid end load: due to the gluing process, the portion of the cantilever beyond $x_\CP$ (gray) is supposed to be infinitely rigid. This rigid end load (colloidal particle, glue, cantilever end) will influence the dynamics of the cantilever because of its inertia in translation and rotation. The center of mass (CoM) of the load is not the bead center anymore, we note $d_\CP$ is its horizontal offset from $x_\CP$.}
\label{fig:sketchRigid} 
\end{figure}

\section{Analytical solution, rigid end load}\label{sec:analytical_solution_Endload}

To take into account the rigidification of the cantilever by the gluing process, it is reasonable to model the portion of the cantilever in contact with the cemented bead as infinitely rigid. The portions of the beam before and after the load can be treated with the Euler-Bernoulli equations, with adequate BC at each end: clamped at $x=0$, free at $x=L$, and criteria matching the inertia in translation and rotation at the load connection. However, one can simplify this complex problem by noting that the portion of the cantilever after the bead is in practical always much shorter than the full cantilever length: $\Delta L \ll L$. This portion will thus be much stiffer, with natural resonance frequencies much higher than those of interest for operation of AFM colloidal probes (see appendix \ref{appendix:modefamilies}). We can therefore forget about the elasticity of this portion of the cantilever, and consider it as part of a rigid end load. In such case, the center of mass of the load is deported along the cantilever axis as well as perpendicular to it, as sketched in Fig.~\ref{fig:sketchRigid}. 

The Euler Bernoulli equation \eqref{eq:EB2} still describes the beam, and clamping conditions \eqref{eq:BC0} in $x=0$ hold. The BC at the last flexible point of the cantilever in $x=x_\CP$ are now:
\begin{subequations}
\label{eq:BCxCP}
\begin{align}
z'' (x_\CP) & = \tilde{m} \alpha^4 \big[\tilde{\rho}^2 z'(x_\CP) + d_\CP z(x_\CP)\big], \label{eq:BCxCP2} \\
z^{(3)} (x_\CP) & = - \tilde{m} \alpha^4 \big[z(x_\CP) + d_\CP z'(x_\CP)\big], \label{eq:BCxCP3} 
\end{align}
\end{subequations}
where $d_\CP$ is the horizontal offset from $x_\CP$ of the center of mass of the rigid load normalised to $L$, and $\tilde{\rho}$ its normalised giration radius. Adapting the previous derivation or following Ref.~\onlinecite{Oguamanam-2003}, we find again eigenmodes shaped as
\begin{equation} \label{eq:psinbar}
\bar{\psi}_n (x) = \psi_c(x) - \bar{R_n} \psi_s(x) \quad \mathrm{for} \quad x\le x_\CP,
\end{equation}
with $\bar{R_n}$ given below, implicitly function of the eigenvalues values $\bar{\alpha}_n(\tilde{m},\tilde{\rho},x_\CP,d_\CP)$, solutions of:
\begin{equation}
\begin{split}
\bar{R_n} &= \frac{\tilde{m}\alpha^4(\tilde{\rho}^2\psi_c'(x_\CP)+d_\CP\psi_c(x_\CP))-\psi_c''(x_\CP)}{\tilde{m}\alpha^4(\tilde{\rho}^2\psi_s'(x_\CP)+d_\CP\psi_s(x_\CP))-\psi_s''(x_\CP)} \\
&= \frac{\tilde{m}\alpha^4(\psi_c(x_\CP)+d_\CP\psi_c'(x_\CP))+\psi_c^{(3)}(x_\CP)}{\tilde{m}\alpha^4(\psi_s(x_\CP)+d_\CP\psi_s'(x_\CP))+\psi_s^{(3)}(x_\CP)}.
\end{split}
\end{equation}
Finally, the normalisation of the eigenmodes $\bar{\phi}_n(x)=\bar{\psi}_n(x)/\sqrt{\langle\bar{\psi}_n^2\rangle}$ is assured with
\begin{align}
\langle\bar{\psi}_n^2\rangle = \int_0^{x_\CP} \bar{\psi}_n^2 (x) \dd x & + \tilde{m} \bar{\psi}_n^2(x_\CP) + \tilde{m} \tilde{\rho}^2 \big[\bar{\psi}_n'(x_\CP)\big]^2 \nonumber\\
& + 2 \tilde{m} d_\CP \bar{\psi}_n(x_\CP)\bar{\psi}_n'(x_\CP).
\end{align}
Note that above $x_\CP$, the normal mode is simply a linear function:
\begin{equation}
\bar{\phi}_n(x) = \bar{\phi}_n(x_\CP) + (x-x_\CP)\bar{\phi}'_n(x_\CP) \quad \mathrm{for} \quad x>x_\CP.
\end{equation}

\section{Application to thermal noise calibration of AFM cantilevers}\label{sec:thermalnoisecalib}

We now explore the application of the computed mode shape to the calibration of AFM cantilevers. The goal is to infer the value of the force $F$ of interaction between the probe and the sample from the signals available in the instrument, in the case of a static loading. $F$ is applied at the contact point between the colloidal probe and the sample, and is supposed to be perpendicular to the sample (no friction).

\begin{figure}[tb]
\begin{center}
\includegraphics{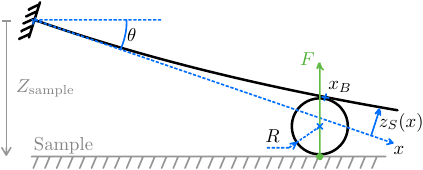}
\end{center}
\caption{Sketch of the colloidal probe configuration during an interaction with a sample. The cantilever is presented at an angle $\theta$ with respect to the horizontal surface, which can be moved vertically along $Z_\mathrm{sample}$. The force $F$ of interaction is supposed to be vertical only (no friction). Since the point of contact with the sample present an horizontal offset from $x_\CP$, $F$ also exerts a moment on the cantilever, responsible for the term in $\tilde{r} \tan\theta$ in Eq.~\ref{EqPhiS} for the static deflection $z_S(x)$.}
\label{FigContact} 
\end{figure}

As sketched in Fig.~\ref{FigContact}, the cantilever is positioned at an angle $\theta$ with respect to the sample. The resulting static deflection is~\cite{Edwards-2008}:
\begin{equation}
z_S(x) = \frac{F}{k_S} \Phi_S(x), \label{EqStaticZ}
\end{equation}
with $k_S=3EI/L^3$ the static stiffness of the bare cantilever, and 
\begin{subequations} \label{EqPhiS}
\begin{align}
\Phi_S(x)&=\frac{1}{2}x^2\left(3x_\CP-3\tilde{r}\tan{\theta}-x\right) \ & \text{for} \ x\le x_\CP,\\
\Phi_S(x)&=\Phi_S(x_\CP)+\Phi_S'(x_\CP)x \ & \text{for} \ x> x_\CP,
\end{align}
\end{subequations}
with $\tilde{r}=R/L$. In most AFMs, the deflection measurement relies on a 4-quadrants photodetector and the Optical Beam Deflection (OBD) technique. The signal is usually a voltage $V$ which, due to the OBD scheme, is proportional to the slope $z'(x_m)$ of the cantilever at the measurement point $x_m$: $V=a z'(x_m)$. The coefficient $a$ is however not available to the user, but $V$ is rather converted to the deflection $Z_m=z_S(x_m)$ using the sensitivity $\sigma$ (in $\SI{}{nm/V}$):
\begin{equation} \label{EqDefSigma}
Z_m = \sigma V.
\end{equation}
Note that $\sigma$ and $a$ are simply linked by the relation $\Phi_S(x_m)=\sigma a \Phi_S'(x_m)$, thus knowing one or the other is equivalent. The calibration of $\sigma$ relies on a force curve on a rigid sample: during the contact, the sample is translated vertically by a calibrated displacement $Z_\mathrm{sample}=\cos \theta z_S(x_\CP)$ while $V$ is measured. The constant compliance part of the curve gives
\begin{equation}
\sigma = \frac{1}{\cos \theta}\frac{\Phi_S(x_m)}{\Phi_S(x_\CP)} \frac{\dd Z_\mathrm{sample}}{\dd V}.
\end{equation}
The AFM user interface usually assumes $x_m=x_\CP$, simplifying the above equation.

From the experimentally calibrated $\sigma$, one can therefore deduce the defection $Z_m$ from the OBD signal $V$. We now need to introduce an effective stiffness $k_\eff$ to interpret the deflection as the static force of interaction: $F=k_\eff Z_m$. From Eq.~\ref{EqStaticZ}, we deduce
\begin{equation} \label{Eqkeff}
k_\eff = \frac{k_S}{\Phi_S(x_m)}.
\end{equation}
We therefore need to estimate $k_S$, the static stiffness of the bare cantilever. Two methods are routinely used thereto: the Sader method or the thermal noise calibration. The former relies on the hydrodynamics of the fluid around the cantilever, which is potentially strongly modified by the colloidal probe, and therefore unfitted. The latter relies on the equipartition theorem which implies that the mean square amplitude $\langle Z_1^2 \rangle$ of the mode 1 deflection in $x=1$ driven by thermal noise is
\begin{equation}
\frac{1}{2}k_B T = \frac{1}{2} k_1 \langle Z_1^2 \rangle,
\end{equation}
with $k_1$ the dynamic stiffness, linked to $k_S$ by~\cite{Laurent-2013}:
\begin{equation} \label{Eqk1}
k_1 = \frac{\alpha_1^4}{3 \phi_1^2(1)} k_S.
\end{equation}

The actual measurement is the voltage $V$, which is interpreted as the measured deflection $Z_m$ through Eq.~\ref{EqDefSigma}, but is set by the slope of the mode 1 during the thermal noise measurement :
\begin{equation}
V = \frac{Z_m}{\sigma} = a z_1'(x_m) = a Z_1 \frac{\phi_1'(x_m)}{\phi_1(1)}.
\end{equation}
The raw stiffness estimated by the instrument is therefore
\begin{equation} \label{Eqk1m}
k_{1m} = \frac{k_B T}{ \langle Z_m^2 \rangle} = \left( \frac{\Phi_S'(x_m) }{\Phi_S(x_m)}\frac{\phi_1(1)}{\phi_1'(x_m)} \right)^2 k_1.
\end{equation}
Combining Eqs.~\ref{Eqkeff}, \ref{Eqk1} and \ref{Eqk1m}, we can define a correction factor $\beta$ to apply to the raw stiffness $k_{1m}$ to deduce the effective static stiffness $k_\eff$:
\begin{align}
k_\eff & = \beta k_{1m},\\ 
\text{with}\qquad \beta & = \frac{3\Phi_S(x_m)}{\alpha_1^4} \left( \frac{\phi_1'(x_m)}{\Phi_S'(x_m)}\right)^2.\label{EqBeta}
\end{align}

The mode shape $\phi_1(x)$ and the eigenvalue $\alpha_1$ depend on the 3 to 4 loading parameters of the cantilever ($\tilde{m},\tilde{\rho},x_\CP$ and $d_\CP$ in the rigid load model). The static deflection depends on $x_\CP$, $\tilde{r}$ and $\theta$, and the measurement point $x_m$ can differ from $x_\CP$. $\beta$ therefore depends on up to seven parameters which change from one probe to another. Our models anyway take fully into accounts those 7 parameters and let one compute $\beta$ in each particular situation. We provide an example in tables \ref{table:punctual} and \ref{table:rigid} of such an evaluation for a generic situation. We consider a typical silicon cantilever for contact mode, $L=\SI{500}{\mu m}$ long, $W=\SI{30}{\mu m}$ wide, $H=\SI{3}{\mu m}$ thick, with a silica bead of radius $R = 0$ to $\SI{100}{\mu m}$, glued at a distance $\Delta L=0$ to $0.3 L$ from the free end. The amount of glue is supposed to be negligible, but for the rigid end load model we assume the last part of the cantilever (from $L-\Delta L$ to $L$) to be rigid. Finally, we suppose that the measurement point $x_m$ is tuned at the bead position $x_\CP$, and that $\theta=\SI{10}{\degree}$.

\begin{table}[!hb]
\caption{Computed values of $\alpha_1$ and $\beta$, generic situation, punctual contact model. As the load is increased, $\alpha_1$, thus $\omega_1$, decreases: the added inertia results in a lower resonance frequency. However, the closer is the load to the base, the less important is the effect, since the effective stiffness at the contact point is increasing. The calibration coefficient $\beta$ is increasing with the load size or when it is closer to the base. It results in a strong correction, up to $\SI{50}{\%}$ for those parameters, due to the mode shape modification by the inertia in translation and rotation, and to the hypothesis that the measurement is taken in $x_m=x_\CP$.}
\begin{center}
\begin{tabular}{|c|c||c|c|c|c|c|c|}
\cline{3-8}
\multicolumn{2}{c|}{\multirow{2}{*}{$\alpha_1$}}& \multicolumn{6}{c|}{$R$ ($\SI{}{\mu m}$)}\\
\cline{3-8}
\multicolumn{2}{c|}{} & 0 & 10 & 20 & 30 & 60 & 100 \\
\hline
\multirow{4}{*}{$x_\CP$}
& 1.0 & 1.875 & 1.715 & 1.289 & 0.990 & 0.594 & 0.398 \\
\cline{2-8}
& 0.9 & 1.875 & 1.751 & 1.366 & 1.063 & 0.640 & 0.427 \\
\cline{2-8}
& 0.8 & 1.875 & 1.783 & 1.450 & 1.147 & 0.696 & 0.462 \\
\cline{2-8}
& 0.7 & 1.875 & 1.811 & 1.539 & 1.246 & 0.763 & 0.504 \\
\cline{2-8}
\hline
\end{tabular}

\vspace{3mm}

\begin{tabular}{|c|c||c|c|c|c|c|c|}
\cline{3-8}
\multicolumn{2}{c|}{\multirow{2}{*}{$\beta$}}& \multicolumn{6}{c|}{$R$ ($\SI{}{\mu m}$)}\\
\cline{3-8}
\multicolumn{2}{c|}{} & 0 & 10 & 20 & 30 & 60 & 100 \\
\hline
\multirow{4}{*}{$x_\CP$}
& 1.0 & 0.817 & 0.885 & 0.984 & 1.022 & 1.084 & 1.177 \\
\cline{2-8}
& 0.9 & 0.906 & 0.940 & 0.998 & 1.030 & 1.098 & 1.206 \\
\cline{2-8}
& 0.8 & 0.999 & 1.011 & 1.026 & 1.045 & 1.117 & 1.243 \\
\cline{2-8}
& 0.7 & 1.085 & 1.089 & 1.071 & 1.070 & 1.143 & 1.293 \\
\cline{2-8}
\hline
\end{tabular}
\end{center}
\label{table:punctual}
\end{table}

The first step is to compute the parameters for the two models. We therefore compute the cantilever mass $m_c = \mu L W H$, with $\mu=\SI{2330}{kg/m^3}$ the density of silicon, and that of the bead $m_\CP= 4 \mu_\CP \pi R^3 / 3$, with $\mu_\CP=\SI{2650}{kg/m^3}$ for silica. For the punctual contact model, we compute as in Ref.~\onlinecite{Laurent-2013}:
\begin{equation}
\begin{split}
\tilde{m} &= \frac{m_\CP}{m_c} \\
\tilde{m}\tilde{\rho}^2 & = \frac{m_\CP}{m_c} \frac{7}{5} \frac{R^2}{L^2}.
\end{split}
\end{equation}
Note that the giration radius $\tilde{\rho}$ is not simply the normalized value of the bead radius $\tilde{r}$, since it also takes into account the inertia of the bead in rotation. For the rigid load model, the last part of the cantilever has to be added to the rigid load, leading to:
\begin{equation}
\begin{split}
\tilde{m} & = \frac{m_\CP}{m_c} + \frac{\Delta L}{L} \\
\tilde{m}\tilde{\rho}^2 & = \frac{m_\CP}{m_c} \frac{7}{5} \frac{R^2}{L^2} + \frac{\Delta L^3}{L^3}\\
\tilde{m}d_\CP & = \frac{1}{2}\frac{\Delta L^2}{L^2}
\end{split}
\end{equation}
Once those parameters evaluated, we can compute numerically the eigenvalue $\alpha_1$ and the associated mode shape $\phi_1$, and finally the correction factor $\beta$. Note that the resonance frequency distribution of all modes, or their shift with respect to an unloaded cantilever, are also available with this approach.

\begin{table}[b]
\caption{Computed values of $\alpha_1$ and $\beta$, generic situation, rigid end load model. $\alpha_1$ and $\beta$ are affected by the load position and inertia in a similar way to Tab.~\ref{table:punctual}, with only small numerical differences. The hypothesis that the lever is rigid beyond $x_\CP$ has little effect even if the attached mass is small, as demonstrated by the first column of $\alpha_1$: the error in $\alpha_1$ with respect to Tab.~\ref{table:punctual} is below $\SI{3}{\%}$, even when rigidifying the last $\SI{30}{\%}$ of the lever.}
\begin{center}
\begin{tabular}{|c|c||c|c|c|c|c|c|}
\cline{3-8}
\multicolumn{2}{c|}{\multirow{2}{*}{$\alpha_1$}}& \multicolumn{6}{c|}{$R$ ($\SI{}{\mu m}$)}\\
\cline{3-8}
\multicolumn{2}{c|}{} & 0 & 10 & 20 & 30 & 60 & 100 \\
\hline
\multirow{4}{*}{$x_\CP$}
& 1.0 & 1.875 & 1.715 & 1.289 & 0.990 & 0.594 & 0.398 \\
\cline{2-8}
& 0.9 & 1.873 & 1.749 & 1.365 & 1.062 & 0.640 & 0.427 \\
\cline{2-8}
& 0.8 & 1.857 & 1.769 & 1.445 & 1.146 & 0.696 & 0.462 \\
\cline{2-8}
& 0.7 & 1.820 & 1.765 & 1.518 & 1.239 & 0.763 & 0.504 \\
\cline{2-8}
\hline
\end{tabular}

\vspace{3mm}

\begin{tabular}{|c|c||c|c|c|c|c|c|}
\cline{3-8}
\multicolumn{2}{c|}{\multirow{2}{*}{$\beta$}}& \multicolumn{6}{c|}{$R$ ($\SI{}{\mu m}$)}\\
\cline{3-8}
\multicolumn{2}{c|}{} & 0 & 10 & 20 & 30 & 60 & 100 \\
\hline
\multirow{4}{*}{$x_\CP$}
& 1.0 & 0.817 & 0.885 & 0.984 & 1.022 & 1.084 & 1.177 \\
\cline{2-8}
& 0.9 & 0.910 & 0.943 & 0.999 & 1.030 & 1.098 & 1.206 \\
\cline{2-8}
& 0.8 & 1.026 & 1.033 & 1.036 & 1.048 & 1.117 & 1.243 \\
\cline{2-8}
& 0.7 & 1.144 & 1.143 & 1.103 & 1.085 & 1.145 & 1.294 \\
\cline{2-8}
\hline
\end{tabular}
\end{center}
\label{table:rigid}
\end{table}

In tables \ref{table:punctual} and \ref{table:rigid}, we report a set of computed values for the generic situation defined above, for the 2 models. Note that in this generic situation, the rigid load model is an approximation of the punctual contact model, and thus monitors to what extent this rigid end approximation is valid. We see that for a small offset $\Delta L$ or a large load, values of $\alpha_1$ and $\beta$ differ very little: in the examples reported here, the difference is lower than 1\% as soon as $x_\CP>0.8$ or $R>\SI{20}{\mu m}$. In these cases indeed, the rigid end approximation is reasonable, as it concerns only a small fraction of the cantilever (small offset), or because this end part has a very little impact on the dynamics compared to the heavy load. Note that the difference on $\beta$ would be more noticeable by measuring in $x_m=1$ instead of $x_m=x_\CP$. Of course, when considering higher order modes, the rigid approximation would fail as well.

The model of our earlier works~\cite{Laurent-2013,Chighizola-2020} corresponds to a bead located at the cantilever end, in $x_\CP=1$, and thus to the first line of both tables. The main novelty of the current work is thus in considering the effect of the load setback, when $x_\CP<1$, reported in the next lines of the tables. Though the effect of the load is less pronounced on the mode shape ($\alpha_1$ is increasing towards the unloaded value) when $x_\CP$ decreases, the effective stiffness raises significantly ($\beta$ can increase by up to $\SI{20}{\%}$ in the parameter range of the table). This is an effect of applying the load closer to the cantilever base, thus shortening the effective length of the cantilever, and increasing its effective stiffness as $1/\Phi_S(x_\CP)\propto x_\CP^{-3}$ (Eq.~\ref{Eqkeff}). The framework developed here allows eventually one to be quantitative on the evaluation of the tip-sample interaction force.

The values of $\alpha_1$ or $\beta$ displayed in tables \ref{table:punctual} and \ref{table:rigid} are specific for the geometry and material of the generic example, and should in general not be used for a particular experiment. As many parameters (up to 7) are needed to describe a colloidal probe and the measurement configuration, reference tables are not handy and we provide in Ref.~\onlinecite{Archambault-Scripts-2023} the scripts to compute directly the values of interest (eigenvalue $\alpha_n$, mode shape $\phi_n$, calibration coefficient $\beta$) for experimentalists using colloidal probes. The value of $\beta=0.817$ for $R=\SI{0}{\mu m}$ and $x_\CP=1$ (no load, measurement at the cantilever tip) is the one commonly found for thermal noise calibrations in commercial AFMs. This coefficient should only be replaced by the updated value from the model to have an accurate calibration of the force measurement.

\section{Experiments}\label{sec:exp}

\begin{figure}
 \centering
 \includegraphics[width=8.5cm]{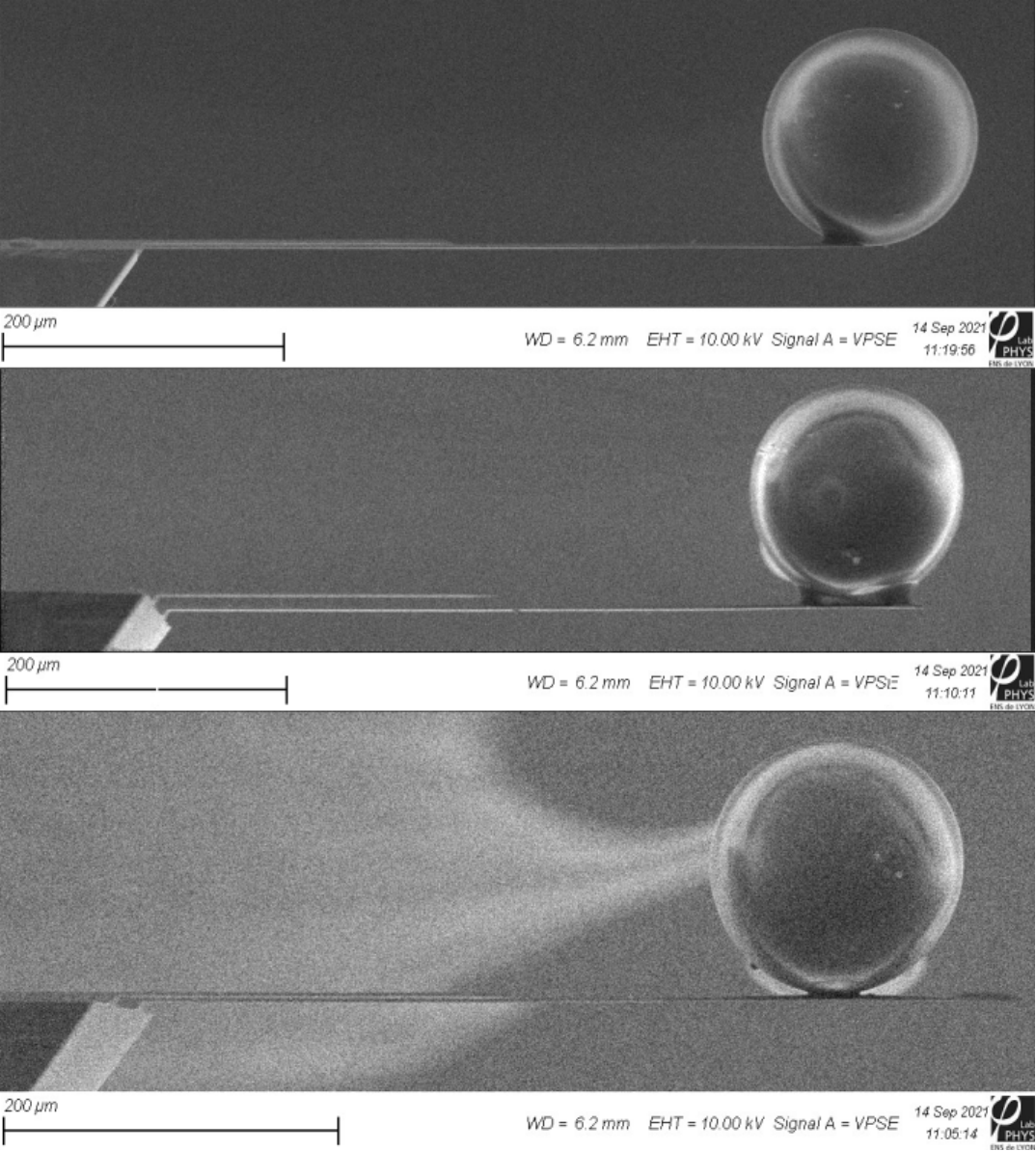}
 \caption{SEM images of the 3 cantilevers with, from top to bottom, $x_\CP = 1$, $x_\CP = 0.9$ and $x_\CP = 0.8$. The trail on the third picture is an artifact from the SEM measurement (insulating polystyrene bead charging under the electron beam).}
 \label{fig:SEM}
\end{figure}

To test our models, we measure along loaded cantilevers the amplitude of the thermal noise driven deflection of the first four flexural modes. The raw samples are uncoated silicon All-In-One type A tipless cantilevers from BudgetSensors, with nominal dimensions $L = \SI{500}{\mu m}$, $H = \SI{2.7}{\mu m}$, and $W = \SI{30}{\mu m}$. They are loaded with polystyrene beads of diameter $2R = \SI{150}{\mu m}$, placed near the tip of the cantilevers using an epoxy glue. Three different samples (respectively A, B and C) have been prepared with different positions of the bead giving values close to respectively $x_\CP = 1$, $0.9$ and $0.8$. The geometry of all cantilevers is characterized using Scanning Electron Microscopy (SEM) (Fig.~\ref{fig:SEM}), except for the thickness which is too small to be precisely measured. It can however be deduced from the measurement of the first resonance frequency of the unloaded cantilever, since Eq.~\ref{eq:dispersion} leads to:
\begin{equation}
H = \sqrt{\frac{12\mu}{E}}\frac{L^2}{\alpha_1^2}\omega_1,
\end{equation}
where all quantities are tabulated ($E=\SI{169}{GPa}$, $\mu=\SI{2330}{kg/m^3}$, $\alpha_1=1.875$) or measured ($L$, $\omega_1$). The measured values of the geometrical parameters and deduced values of the dimensionless parameters $\tilde{\rho}$ and $\tilde{m}$ are presented in appendix, table \ref{tab:SEM_measurements}.

\begin{figure}[htb]
\includegraphics{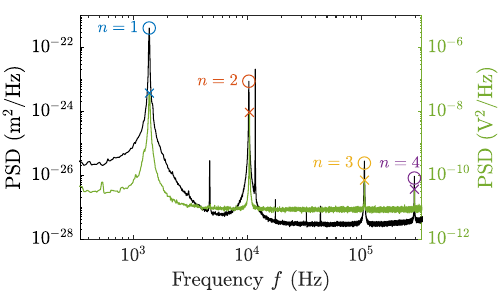}
\caption{\label{ExpSpectrum} Power spectrum density of the thermal noise driven deflection of cantilever B loaded with a polystyrene bead. The different peaks correspond to the different flexural (labelled with $n$) and torsional (unlabelled) modes of the cantilever. The black PSD (left scale, units $\SI{}{m^2/Hz}$, $\circ$ markers for peaks) is measured at $x = 0.68$ (near an anti-node of mode 4) with an interferometer~\cite{Paolino-2013}, while the green PSD (right scale, units $\SI{}{V^2/Hz}$, $\times$ markers for peaks) is measured around $x=0.8$ with a commercial AFM. Amplitudes cannot be compared (true deflection versus uncalibrated slope, different locations), but the 4 resonance frequencies can be precisely extracted from both spectra, regardless of the sensing position $x$ or instrument.}
\end{figure}

\subsection{Interferometric measurement}

The power spectrum density (PSD) $S_d(f)$ of the deflection is measured on a large frequency bandwidth for the 3 cantilevers using a quadrature phase differential interferometer~\cite{Paolino-2013} featuring a high spectral resolution reaching $\SI{e-14}{m/\sqrt{Hz}}$. This interferometric setup gives access to the calibrated vertical deflection $z(x,t)$ along cantilevers, where 4-quadrant photodiode setups measure the slope of this deflection. Fig.~\ref{ExpSpectrum} shows an example of a measured PSD. Different peaks can be seen, corresponding to the different flexural or torsional modes of the cantilever, driven by thermal noise only. The resonance frequency of the first 4 flexural modes, both for the raw ($f_n^r$) and loaded ($f_n^l$) samples, are extracted from those spectra, and reported in appendix, table \ref{tab:f_n}. The relative uncertainty on those measurements is very small, typically below $10^{-3}$, making them an excellent marker to track the effect of the loading.

The resonance frequencies $f_n$ are linked to the spatial eigenvalues $\alpha_n$ by Eq.~\ref{eq:dispersion}, with a multiplicative factor that is unchanged when loading the cantilever. We therefore expect the ratio of loaded (superscript $^l$) to unloaded (superscript $^r$) values of the resonance frequency to be \begin{equation}
 \frac{f_n^l}{f_n^r}=\left(\frac{\alpha_n^l(\tilde{m},\tilde{\rho},x_\CP,d_\CP)}{\alpha_n^r}\right)^2
\end{equation}
where the dependency of $\alpha_n^l$ in $d_\CP$ only stands if we consider the end load model. Since we can compute this ratio for four modes in our experiment, we have a set of 4 equations with 3 unknowns ($\tilde{m},\tilde{\rho},x_\CP$) for the single point contact model, and a fourth one ($d_\CP$) for the rigid end load model. We can therefore extract the values of those 3 or 4 parameters with a good precision using the frequency shifts from raw to loaded. For the single point contact model, we can even extract those parameters from the loaded frequencies alone: studying the ratio $f_n^l/f_1^l$ with $n=2$ to 4 leads to 3 equations with 3 unknowns. This approach relying on the measurement of resonance frequencies of the loaded cantilever alone could be expanded to the rigid end load model by considering more modes (5 at least). The values extracted for the 3 samples and the different models and approach are reported in the appendix tables \ref{tab:mrxB} to \ref{tab:mrxB3}, they are in reasonable agreement to what is expected from the geometry. As those values are extracted from measurements, they can be more accurate as they include the effect of the glue, defects on the cantilever or bead, contamination, deviations from the tabulated densities, etc. Note that when extracting the model parameter values, some precautions need to be taken with respect to the mode order, as detailed in appendix \ref{appendix:modefamilies}.

\begin{figure}[!tb]
\null\vspace{8mm}

\includegraphics{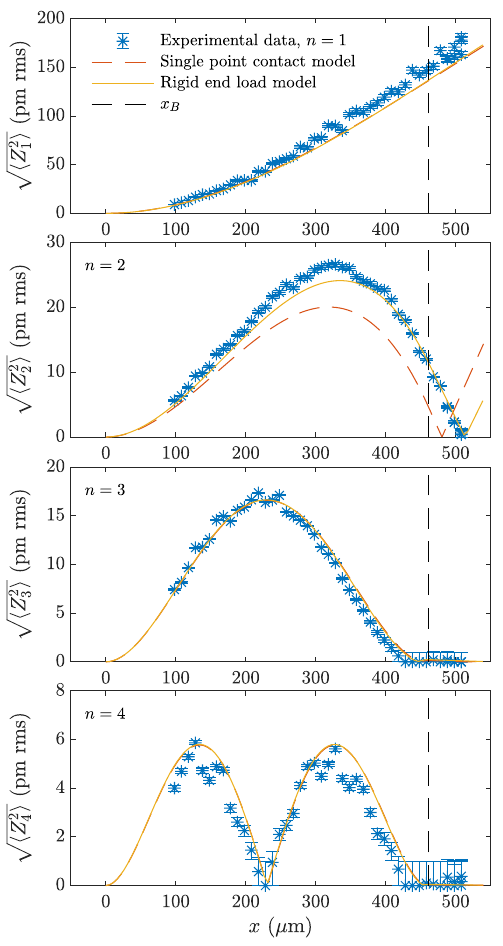}
\caption{\label{FitModes} Thermal noise amplitude for the different modes measured along the cantilever (blue stars) for cantilever B ($x_\CP \sim 0.9$, materialized on the graph with the vertical black dashed line). The error bars correspond to the statistical uncertainty and floor noise of the measurement process. The result of a simultaneous fit on the four modes using the point contact model (dashed red) and the rigid end load model (plain line, orange) is superposed. Since the amplitude of the mode is measured as $\sqrt{Z_n^2}=[\int S^2(\omega) d\omega]^{1/2}$, the absolute value of each mode is plotted for the fitted model. The large load results in a strong inertia in translation and rotation at the cantilever end, especially for higher order modes since inertial forces scale as $\omega^2$. This is equivalent to an effective clamping boundary condition, an effect clearly visible on modes 3 and 4: end position and slope are both zero.}
\end{figure}

The PSD of thermal noise can be measured at different positions $x$ along the cantilever, revealing the modes' shape~\cite{Paolino-2009,Laurent-2013}: by integrating the spectrum on a small frequency window around the different peaks, we construct the spatial profile for each mode, as illustrated in Fig.~\ref{FitModes}. We can superpose on those profiles the expectations from the two models: the only parameters left to determine are the length $L$ and origin $x_0$ of the position measurement, and the stiffness $k_S$ of the cantilever that drives the amplitude of the thermal noise. We perform a simultaneous fit of the four modes profiles to extract those parameters and plot the expectation from the models in Fig.~\ref{FitModes} for cantilever B, and in appendix for the two other samples (Fig.~\ref{FitModesAC}). The two models provide a very good description of the mode shape and thermal noise amplitude, except for the single contact point model for mode 2. This small deviation between the model and the measurement could either come from an excess extrinsic noise superposing to the thermal fluctuations in this frequency range, or from the model itself which is oversimplifying the actual physical system. The stiffnesses deduced from the fits are anyway consistent with the ones measured on the raw samples, as one would expect. For cantilever B presented in Fig.~\ref{FitModes} for example, we estimate $k_S=\SI{0.14}{N/m}$ for all models, before or after gluing the bead.

\subsection{Commercial AFM measurement}

As most commercial AFMs use the OBD scheme to measure the deflection of the cantilever, the experimental study of previous section cannot be applied directly. Scanning the cantilever along its length is anyway time-consuming and not a calibration step one would like to perform routinely: the purpose of this measurement was to demonstrate the pertinency of the model to describe the shape of the resonance modes. We report in Fig.~\ref{ExpSpectrum} the thermal noise spectra acquired on cantilever B with a commercial AFM (JPK Nanowizard 4). The four resonance peaks are clearly resolved, leading to the same frequencies that were used in the previous analysis. This single point measurement is fast and easy to perform, and gives as before the load parameters $\tilde{m},\tilde{\rho},x_\CP,\ldots$ by studying the frequency shift from raw to loaded cantilever, or the frequency ratio between the modes of the loaded cantilever.

Once $x_\CP$ is precisely known, one can set the measurement laser position $x_m$ in $x_\CP$, and perform a classic calibration of the apparatus. First a force curve on a hard surface is performed to measure the OBD sensitivity $\sigma$ in $\SI{}{nm/V}$. With the large beads used in this study, a significant friction force leads to an hysteresis in the force curve between the approach and retract segments, and $\sigma$ is measured using the mean of the two slopes to cancel the friction effect~\cite{Stiernstedt-2005}, leading for example to $\sigma=\SI{80}{nm/V}$ for cantilever B. Then a thermal noise measurement is performed to extract the thermal noise amplitude $\langle Z_{1m}^2 \rangle$ (area under the first peak in the spectrum in $\SI{}{V^2/Hz}$ times $\sigma^2$), from which the measured stiffness $k_{1m}$ is estimated. The coefficient $\beta$ computed within our framework from the load parameters using Eq.~\ref{EqBeta} can then be applied to compute the effective stiffness of the cantilever. In this case on the example of cantilever B, the measured value of the dynamic stiffness is $k_{1m}=\SI{0.26}{N/m}$. The stiffness calibrated using the default settings of the AFM (using $\beta=0.817$) would then be $\SI{0.21}{N/m}$, while we compute here $\beta=1.184$ and thus an actual effective stiffness $k_\eff=\SI{0.30}{N/m}$, $\SI{50}{\%}$ higher than the default estimation! An intermediate step of the computation also leads to the static stiffness $k_S=(0.18\pm0.03)\SI{}{N/m}$, in reasonable agreement with the interferometric measurement.

\section{Conclusions}

In this article, we present two analytical approaches to deal with functionalized cantilevers of uniform rectangular cross section and high aspect ratio $L/W\gg 1$ with a load attached anywhere along its length. The simple point contact model takes into account the inertia in translation and rotation added by the load in a single point of the cantilever, which is allowed to deform all along its length. This model is expected to be pertinent for geometrically small loads or for small contact areas between the load and the cantilever. The rigid end load model simply builds upon the hypothesis that the cantilever is very rigid beyond the contact with the load, and should be valid for large loads close to the extremity, or large amounts of glue that rigidify the end part of the lever. We solve the Euler-Bernoulli equations corresponding to these cases and study the corresponding mode shapes. This leads to calibration coefficients that can be used to interpret thermal noise measurement in standard AFMs, and deduce the probe static stiffness. Finally, we performed some measurements on 3 different samples to illustrate the approach. From thermal noise spectra, we measure the resonance frequencies of the first modes of the cantilever, from which we deduce the properties of the load: mass, gyration radius, position. A comparison between the mode shapes, analytically computed, and extracted from the thermal noise measurements, demonstrates the relevance of the approach. For the large loads we probe here (beads of diameter $\SI{150}{\mu m}$ for cantilevers $\SI{500}{\mu m}$ long), both models work, though the rigid end load model leads to slightly better results.

The present work is limited to rectangular cantilevers, and future work could address the case of triangular probes, which are also common in AFM. It can be noted however that if only the end is triangular on a short length, with most of the cantilever length rectangular of uniform cross section, then the rigid end load model can be used as an elegant way to deal with the non rectangular end of the probe. Indeed in such case, for the first oscillation modes, the triangular tip can be considered rigid, and an end load to the straight part of the cantilever. Another limitation of the present work is the description of the normal modes within the Euler-Bernoulli framework. While pertinent for high aspect ratio cantilevers, it should be considered with care for wide cantilevers. In particular, high order eigenmodes with a spatial wavelength comparable to or smaller than $W$ will deviate from the current framework, with 2D effects in the transverse direction.

A key learning from this study is that the knowledge of several resonance frequencies of the cantilever, ideally before and after loading, but potentially of the loaded cantilever alone, can be enough to extract the load parameters (mass, gyration radius, position). Indeed, the ratio of these frequencies delivers this information regardless of the cantilever geometry, and is eventually enough to conclude on the probe parameters. Introducing the angle $\theta$ made by the cantilever and the sample, and the normalized radius of the colloidal probe $\tilde{r}$, the calibration coefficient $\beta$ can be computed, and substituted to the default one used in most instruments. This procedure allows to decrease the systematic calibration errors due to the geometry in order to make quantitative force measurements with any AFM. For practical purposes, functional scripts to compute the modes shapes, eigenvalues, and calibration coefficient $\beta$ are available in Ref.~\onlinecite{Archambault-Scripts-2023}.

\begin{acknowledgments}
We thank A. Podestà, A. Petrosyan, V. Dolique and S. Ciliberto for enlightening technical and scientific discussions. Part of this work has been financially supported by the Agence Nationale de la Recherche through grants ANR-18-CE30-0013 and ANR-18-CE08-0023.
\end{acknowledgments}

\appendix

\section*{Appendix}

\subsection{Data availability}
The data that support the findings of this study are available on Zenodo at doi:10.5281/zenodo.10102664~\cite{Archambault-Dataset-2023}. The scripts to compute the modes shapes, eigenvalues, and calibration coefficient $\beta$ are available in Ref.~\onlinecite{Archambault-Scripts-2023}.

\subsection{Cantilevers' geometrical characteristics}

\begin{table}[thb]
\caption{Geometrical parameters measured from SEM images. The cross section of the cantilevers is actually trapezoidal, and characterized by its large width $W_\mathrm{top}$ and small one $W_\mathrm{bottom}$. The thickness $H$ is deduced from the first resonance frequency of the raw cantilever. The stiffness deduced from the geometry $k_S^\mathrm{Geo}$ matches the one deduced from the thermal noise calibration~\cite{Butt-1995} of the raw cantilever $k_S^r$.}
 \centering
 \begin{tabular}{|l|c|c|c|}
 \hline
 Sample & A & B & C \\
 \hline
 $L$ ($\mu$m) & $499 \pm 12$ & $523 \pm 12$ & $518 \pm 12$ \\
$W_\mathrm{top}$ ($\mu$m) & $27.7 \pm 1.4$ & $35.3 \pm 0.2$ & $35.4 \pm 0.2$ \\
$W_\mathrm{bottom}$ ($\mu$m) & $14.1 \pm 1.4$ & $24.1 \pm 0.2$ & $23.9 \pm 0.2$ \\
$H$ ($\mu$m) & $2.93 \pm 0.14$ & $2.42 \pm 0.11$ & $2.52 \pm 0.12$ \\
$R$ ($\mu$m) & $75 \pm 2.5$ & $75 \pm 2.5$ & $75 \pm 2.5$ \\
$\tilde{m}$ & $26.1 \pm 3.4$ & $21.2 \pm 2.6$ & $20.6 \pm 2.5$ \\
$\tilde{\rho}$ & $0.18 \pm 0.01$ & $0.17 \pm 0.01$ & $0.17 \pm 0.01$ \\
$x_\CP$ & $1 \pm 0.05$ & $0.9 \pm 0.05$ & $0.81 \pm 0.05$ \\
$k_S^\mathrm{Geo}$ (N$/$m) & $0.176 \pm 0.015$ & $0.125 \pm 0.008$ & $0.146 \pm 0.010$ \\
$k_S^r$ (N$/$m) & $0.177 \pm 0.003$ & $0.141 \pm 0.002$ & $0.137 \pm 0.019$ \\ \hline
 \end{tabular}
 \label{tab:SEM_measurements}
\end{table}

Table \ref{tab:SEM_measurements} reports the geometrical dimensions of the cantilever (widths $W$ and length $L$) and of the glued bead (radius $R$ and position $x_\CP$), measured on SEM images. The cantilever thickness $H$ is deduced from the first resonance frequency and the Euler-Bernoulli description of the cantilever. The dimensionless mass $\tilde{m}$ and gyration radius $\tilde{\rho}$ are computed from those measurements, using $\mu=\SI{2340}{kg/m^3}$ for the density of silicon and $\SI{1050}{kg/m^3}$ for polystyrene. The stiffness computed from the geometry (using $E=\SI{169}{Gpa}$ for the Young modulus of silicon in the 110 orientation corresponding to the cantilever long axis) is a good agreement with the stiffness measured from the thermal noise of the unloaded sample.

\subsection{Raw and loaded cantilevers' resonance frequencies}

Table \ref{tab:f_n} reports the resonance frequencies of the raw and loaded samples, read from the thermal noise PSD.

\begin{table}[htb]
\caption{Resonance frequency $f_n$ of the first four modes ($n=1$ to $4$) for the raw (superscript $^r$) and loaded (superscript $^l$) cantilevers.}
 \centering
 \begin{tabular}{l|l|c|c|c|}
 \cline{2-5}
 & Sample & A & B & C \\
\hline
\multicolumn{1}{|l|}{\multirow{2}{*}{$n=1$}} & $f_1^r$ (kHz) & $15.80 \pm 0.05$ & $12.02 \pm 0.04$ & $12.76 \pm 0.04$ \\
\multicolumn{1}{|l|}{} & $f_1^l$ (Hz) & $1420.1 \pm 1.5$ & $1370.3 \pm 1.5$ & $1716.7 \pm 1.7$ \\
\hline
\multicolumn{1}{|l|}{\multirow{2}{*}{$n=2$}} & $f_2^r$ (kHz) & $99.19 \pm 0.17$ & $75.88 \pm 0.12$ & $79.64 \pm 0.13$ \\
\multicolumn{1}{|l|}{} & $f_2^l$ (Hz) & $11007 \pm 6$ & $10305 \pm 5$ & $10378 \pm 5$ \\
\hline
\multicolumn{1}{|l|}{\multirow{2}{*}{$n=3$}} & $f_3^r$ (kHz) & $277.4 \pm 0.5$ & $213.4 \pm 0.3$ & $221.9 \pm 0.3$ \\
\multicolumn{1}{|l|}{} & $f_3^l$ (kHz) & $107.1 \pm 1.7$ & $106.55 \pm 0.04$ & $132.29 \pm 0.05$ \\
\hline
\multicolumn{1}{|l|}{\multirow{2}{*}{$n=4$}} & $f_4^r$ (kHz) & $542.6 \pm 0.8$ & $419.7 \pm 0.5$ & $431.8 \pm 0.5$ \\
\multicolumn{1}{|l|}{} & $f_4^l$ (kHz) & $293.64 \pm 0.07$ & $292.67 \pm 0.07$ & $363.59 \pm 0.09$ \\
\hline
\end{tabular}
\label{tab:f_n}
\end{table}

\subsection{Thermal noise profiles and mode shape fits}

\begin{figure*}[tb]
\includegraphics{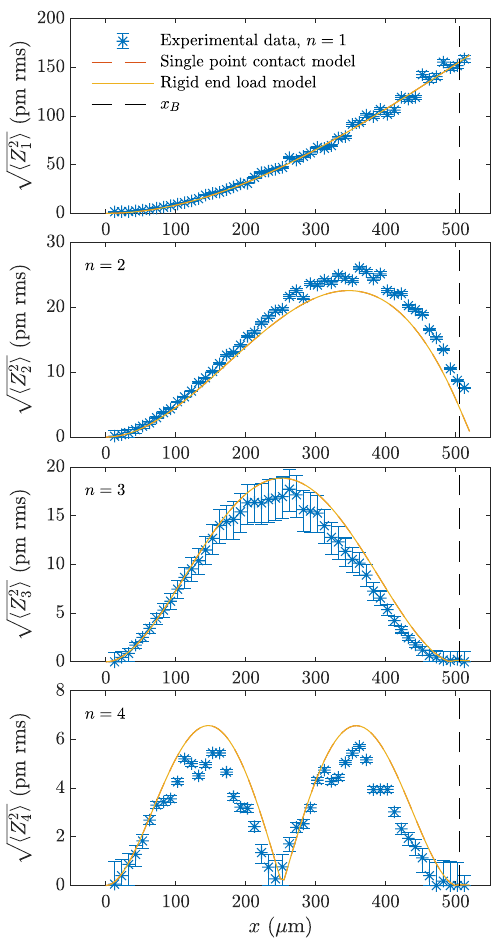} \hfill \includegraphics{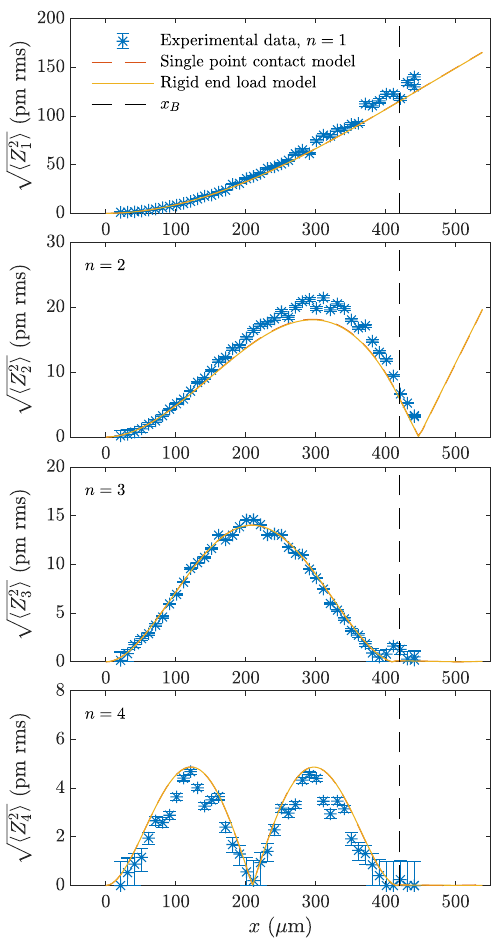}
\caption{\label{FitModesAC} Same as Fig.~\ref{FitModes} for cantilever A ($x_B \sim 1$, left) and C ($x_B \sim 0.8$, right).}
\end{figure*}

Fig.~\ref{FitModesAC} reports the thermal noise profile measured on cantilevers A and C, with the result of the two models superposed. Both describe accurately the experimental data, and are actually almost equivalent. The best fit parameters are reported in Tables \ref{tab:mrxB} to \ref{tab:mrxB3}.

\begin{table}[b]
\caption{Best fit parameters, single point contact model, with $\tilde{m}$, $\tilde{\rho}$ and $x_\CP$ extracted from the resonance frequency shift between raw and loaded samples, while $k_S$ is fitted from the thermal profiles of Figs.~\ref{FitModes} and \ref{FitModesAC}. Estimated uncertainty is not reported when below the last displayed digit.}
 \centering
 \begin{tabular}{|l|c|c|c|}
 \hline
 Sample & A & B & C \\
 \hline
$\tilde{m}$ & $30.8 \pm 0.2$ & $27.7 \pm 0.2$ & $25.3 \pm 0.2$ \\
$\tilde{\rho}$ & $0.152$ & $0.140$ & $0.164$ \\
$x_\CP$ & $0.972$ & $0.856$ & $0.780$ \\
$k_S$ (N$/$m) & $0.154 \pm 0.004$ & $0.137 \pm 0.002$ & $0.145 \pm 0.002$ \\
 \hline
 \end{tabular}
 \label{tab:mrxB}
\end{table}

\begin{table}[b]
\caption{Best fit parameters, single point contact model, with $\tilde{m}$, $\tilde{\rho}$ and $x_\CP$ extracted from the resonance frequency ratio of the loaded samples, while $k_S$ is fitted from the thermal profiles of Figs.~\ref{FitModes} and \ref{FitModesAC}. Those best fit parameters are close to the ones of Tab.~\ref{tab:mrxB}, showing the equivalence of both approaches. Estimated uncertainty is not reported when below the last displayed digit. *Note that for sample B, the fit is not converging unless we fix the value of $x_\CP$. }
 \centering
 \begin{tabular}{|l|c|c|c|}
 \hline
 Sample & A & B & C \\
 \hline
$\tilde{m}$ & $31.5 \pm 5.2$ & $28.6$ & $25.4$ \\
$\tilde{\rho}$ & $0.156 \pm 0.025$ & $0.139$ & $0.170$ \\
$x_\CP$ & $0.997 \pm 0.160$ & $0.856$* & $0.807$ \\
$k_S$ (N$/$m) & $0.17 \pm 0.03$ & $0.137 \pm 0.002$ & $0.160 \pm 0.002$ \\
 \hline
 \end{tabular}
 \label{tab:mrxB2}
\end{table}

\begin{table}[htb]
\caption{Best fit parameters, rigid end load model, with $\tilde{m}$, $\tilde{\rho}$, $x_\CP$ and $d_\CP$ extracted from the resonance frequency shift between raw and loaded samples, while $k_S$ is fitted from the thermal profiles of Figs.~\ref{FitModes} and \ref{FitModesAC}.}
 \centering
 \begin{tabular}{|l|c|c|c|}
 \hline
 Sample & A & B & C \\
 \hline
$\tilde{m}$ & $30.8 \pm 0.2$ & $23.0 \pm 5.5$ & $25.7 \pm 0.2$ \\
$\tilde{\rho}$ & $0.152$ & $0.18 \pm 0.06$ & $0.163$ \\
$x_\CP$ & $0.972$ & $0.856$ & $0.780$ \\
$d_\CP$ & $0$ & $0.05 \pm 0.07$ & $0$ \\
$k_S$ (N$/$m) & $0.154 \pm 0.004$ & $0.135 \pm 0.005$ & $0.145 \pm 0.002$ \\
 \hline
 \end{tabular}
 \label{tab:mrxB3}
\end{table}

\subsection{Single point contact model: families of modes} \label{appendix:modefamilies}

\begin{figure}[htb]
\includegraphics{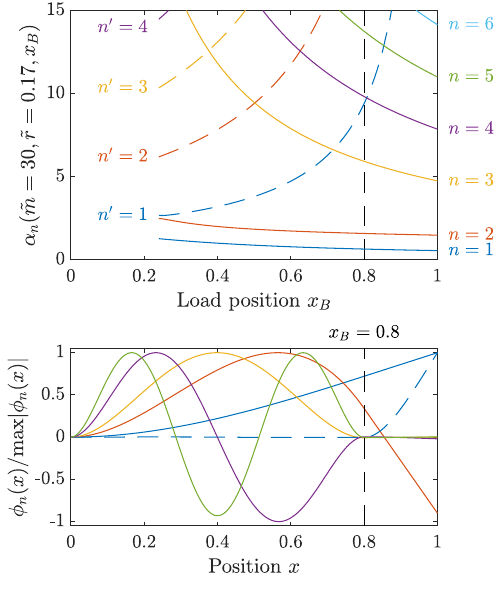}
\caption{\label{fig:modefamilies} (Top) 6 lowest values of $\alpha_n(\tilde{m}, \tilde{\rho}, x_\CP)$ computed for the single point contact model with $\tilde{m}=30$, $\tilde{\rho}=0.17$, and $x_\CP$ from $0.25$ to $1$. (Bottom) Mode shapes corresponding to $x_\CP=0.8$, using the same color code as the top figure to identify the modes.}
\end{figure}

In Fig.~\ref{fig:modefamilies}, we plot the 6 lowest values of $\alpha_n(\tilde{m}, \tilde{\rho}, x_\CP)$ computed for the single point contact model with $\tilde{m}=30$, $\tilde{\rho}=0.17$ as a function of the load position $x_\CP$. We notice very different behaviors: one family is increasing with $x_\CP$, while another is decreasing. If we look at the corresponding mode shape $\phi_n(x)$, we see that the ``decreasing'' modes (labelled with $n$) are similar to the unloaded ones, with the load acting as a fixed position for $n>1$, due its large inertia. The other family (labelled with $n'$) on the contrary corresponds to normal modes of the part of the cantilever beyond the load, and present increasing resonance frequencies when its effective length $L(1-x_\CP)$ shortens. When fitting the model parameters $x_\CP$ explores a small range where modes can change their order between families, so one should be careful to always track the right family. In this article, we restrict the fit to family mode $n$, as the family $n'$ is not experimentally observed: we have few clean measurements beyond $x_\CP$, since the glue often degrades the reflectivity of the cantilever in this area. It implies than when $x_\CP$ can get close to $0.8$ (the case of samples B and C), we need to make sure not to mistake the mode $n=4$ and $n'=1$, since they swap their order in the eigenvalue list. A simple criterion of the position of the maximum of $\phi_n(x)$ with respect to $x_\CP$ works very well in this case.

\bibliography{LoadedCantilever.bib}

\end{document}